# Computational compressive imaging reconstruction with a simulated-annealing-based Ising solver

Ze Zheng [1,2], Xiaoyan Wu[1], Hang Xu [1,2], Jingzheng Huang[1,2,3], Tailong Xiao[1,2,3*], and Guihua Zeng[1,2,3]

[1] Institute for Quantum Sensing and Information Processing, State Key Laboratory of Photonics and Communications, Shanghai Jiao Tong University, Shanghai 200240, China
[2]Hefei National Laboratory, Hefei 230088, China
[3]Shanghai Research Center for Quantum Sciences, Shanghai 201315, China

[*]Corresponding author. Email: tailong_shaw@sjtu.edu.cn

Computational compressive imaging, such as single-pixel imaging and lensless imaging, is a promising imaging technique, enabling the reconstruction of objects from under-sampled measurements through compressed sensing algorithms or deep learning methods. However, the demand for electronic computational resources grows exponentially when recovering high-resolution/multi-dimensional images and training neural networks with numerous parameters, which would become a bottleneck of this technique for widespread adoption. Here, we propose a physics-driven reconstruction framework, which formulates the imaging inverse problems as Ising models and directly recovers target objects through the highly efficient self-evolution toward the minimum Hamiltonian via non-von Neumann Ising machines. We employ a simulated-annealing-based Ising solver to demonstrate this proof-of-concept by reconstructing the objects for experiments of both single-pixel imaging and a lensless diffuser camera at low-sampling rates (ranging from 1 % to 25 %). By further combining the foreground-sparsity constraint and the smoothness constraint, the reconstruction quality of our method is comparable to or even better than that of the mainstream algorithms. This approach has the potential to drastically reduce the electronic computational burden of inverse reconstruction, paving the way for real-time, low-power computational imaging frameworks.

## Introduction

Computational compressive imaging recovers an object from far fewer measurements than the pixel number of the image to be reconstructed, which shifts complexity from the optical front-end to a computational back-end [1-3]. In single-pixel imaging, a sequence of structured patterns illuminates the scene and the resulting intensities are integrated by a single bucket detector, and the image is reconstructed from the correlation between the patterns and the measurements [4-14]. In lensless imaging, such as coded-aperture imaging [15-20] and diffuser camera [21-26], the imaging lens is replaced by a custom mask or a scattering medium, and the sensor captures a multiplexed measurement that contains the multi-dimensional information of the scene. The image is computationally reconstructed from the correlation between the point spread functions (PSFs) and the under-sampled measurement [27, 28]. The forward model [29] is used to describe both optical encoding processes, and the imaging reconstruction is an ill-posed inverse problem as the number of measurements is smaller than the pixel number of the target image. Compressed sensing (CS) algorithms [4, 5] and deep learning methods [30-33] are employed to invert the forward process, however, the required memory, floating-point operations, and energy of both routes grow rapidly with image resolution and dimensionality [34] based on von Neumann electronic computers. CS algorithms typically require thousands of forward and adjoint operations to recover a medium-quality image at medium-sampling rates. Data-driven models should be trained on large datasets with millions of parameters and retrained whenever the optical encoding changes. This back-end cost has become a practical bottleneck that limits the deployment of compressive imaging in real-time, low-power, and embedded scenarios.

Ising machine (IM) offers a physical hardware solution scheme to combinatorial optimization problems, which has shown orders-of-magnitude advantages in time-to-solution [35-37] or energy-to-solution [38-40] than traditional von Neumann schemes for specific problems, such as Max-cut problems [41] and traffic optimization problems [42]. The Ising Hamiltonian is defined as $H_{Ising}(s) = -\sum_{1\le i<j\le N} J_{ij} s_i s_j - \sum_{i=1}^{N} h_i s_i$ with binary spins $s_i = \pm 1$, where i and j are element index, and N is the spin number, which is a universal target for any problem in the complexity class nondeterministic polynomial time (NP) problems [43]. The solution of the original problem corresponds to the lowest-energy spin configuration. Ising machines embody the spin system in physical hardware and let it evolve toward the ground state, thereby circumventing the von Neumann bottleneck of continuously shuttling data between memory and processor [44]. Existing Ising machine implementations span superconducting-circuit quantum annealers [45], coherent Ising machines (CIMs) built from networks of degenerate optical parametric oscillators [46], spatial-photonic Ising machines (SPIMs) that encode thousands of spins into a modulated wavefront [47-50], as well as spintronic, memristive, and CMOS annealing architectures [51]. Ising machines have recently been applied in solving L0-regularized compressed sensing [52, 53] and tomographic reconstruction problems [54]. Nevertheless, these demonstrations remain largely restricted to synthetic one-dimensional sparse signals or magnetic-resonance data on a small scale, and a general Ising reconstruction framework of computational compressive imaging systems has not yet been established.

In this paper, we propose a computational compressive imaging reconstruction framework driven by an Ising machine, which would assist in real-time, energy-efficient computational imaging. We demonstrate that the imaging inverse reconstruction problem, as well as prior constraints, can be formulated as an Ising model, followed by uploading into the Ising machine to evolve to the lowest-energy spin configuration. The reconstructed image is then retrieved from this spin configuration. We employ a simulated-annealing-based Ising solver to perform the evolution, which is the common strategy of SPIMs [47-50] and baseline competitors from other Ising machines [35-37], due to the available Ising machines being unable to meet the requirement of bit resolution of our applications [55-57]. It should be note that various developing Ising machines exhibit solution quality better than simulated-annealing-based algorithms, alongside reduced time consumption and energy efficiency, for solving combinatorial optimization problems. This underpins our rationale that the proposed Ising reconstruction framework can possess practical advantages with advanced Ising machines, provided the coding requirement of bit resolution is satisfied [35-42]. We demonstrate the reconstruction of experiments of both single-pixel imaging and a diffuser camera, and the peak signal-to-noise ratios (PSNRs) and structural similarity index measures (SSIMs) are comparable to or better than the mainstream algorithms. Our proposed framework is also suitable for other computational imaging techniques incorporating multiple imaging modalities, and would further inspire the development of intelligent machine vision.

**Methods**

As shown in Fig. 1(a), in computational imaging systems, the forward model is usually used to describe the optical front-end mapping from the object field to the measured signals, as below:

$$y = Ax, \tag{1}$$

where $y \in \mathbb{R}^{M\times 1}$ is the measurement vector, $A \in \mathbb{R}^{M\times N}$ is the measurement matrix, and $x \in \mathbb{R}^{N\times 1}$ is the vectorized image of the object. M is the number of measurements, and N is the pixel number of the reconstructed image. In single-pixel imaging, each row of $A$ represents an illumination pattern. In the diffuser camera, $A$ corresponds to the linear transmission matrix of the diffuser. Within the range of the optical memory effect, $A$ is a circulant matrix constructed from the PSF. In computational compressive imaging, the number of measurements M is fewer than the image's pixel number N, and Eq. (1) becomes an underdetermined system of equations and admits no unique exact solution. To solve this ill-posed problem, we take the minimization of the $\ell_2$-norm as our optimization objective to minimize the data fidelity loss, as shown in Eq. (2):

$$\hat{x} = argmin_{x\geq 0}\|y - Ax\|_2^2 \,. \tag{2}$$

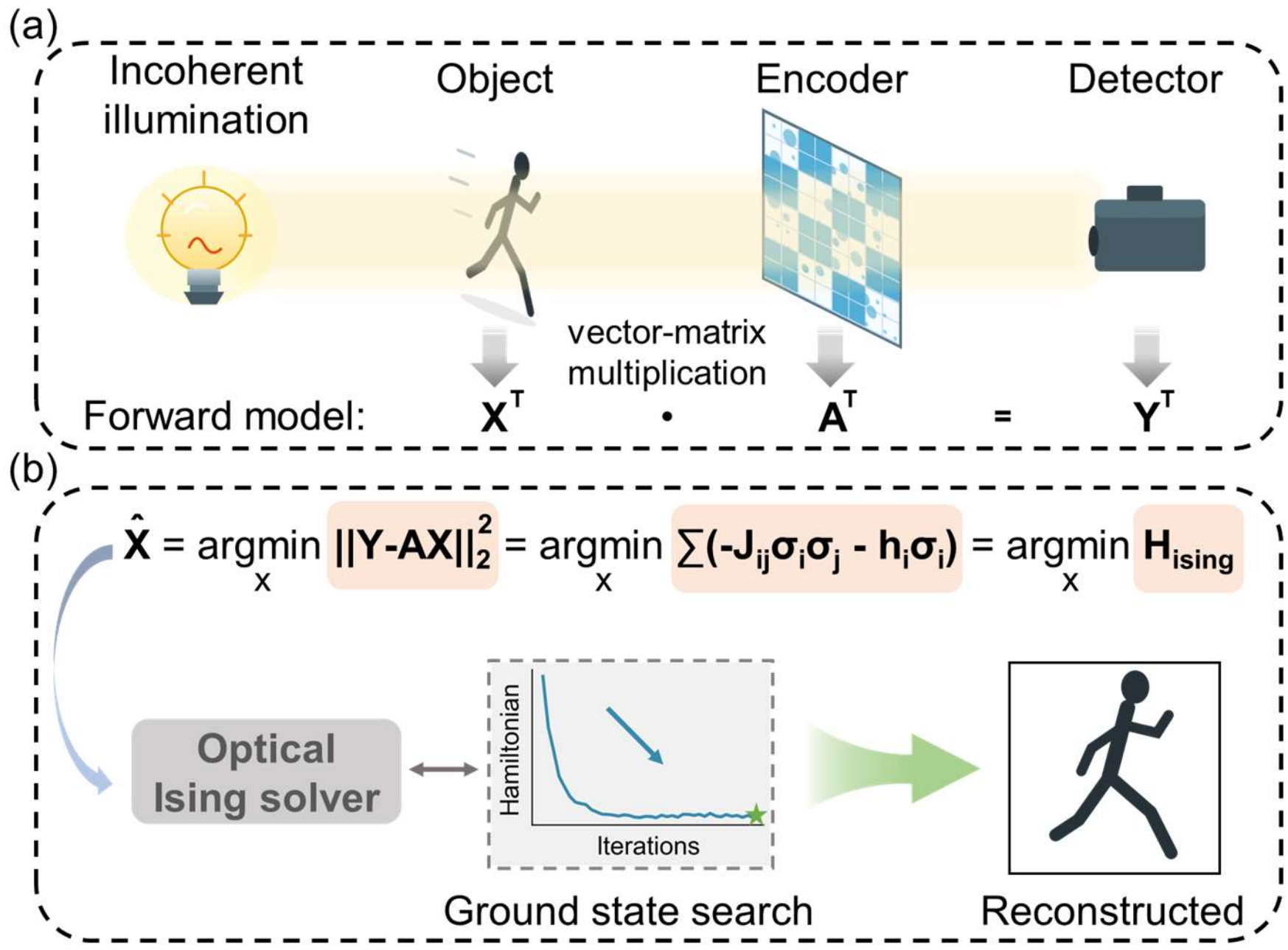

**Figure 1.** Concept and principle of the proposed imaging reconstruction framework driven by an Ising machine. (a) The general optical front-end encoding process, which is described by the forward model. The object *X* is illuminated by incoherent light, followed by encoding by an optical modulator. *A* represents the measurement matrix. The detector measures the one- or two-dimensional intensity *Y* of the resulting light field corresponding to different imaging techniques. (b) The reconstruction framework using the Ising machine. The image to be reconstructed, represented by minimizing the $\ell_2$-norm, can be further modeled as the minimization of the Ising Hamiltonian, followed by uploading into the Ising solver for self-evolution. The reconstructed image can be recovered directly by reshaping the spin configuration corresponding to the Ising minimum Hamiltonian (i.e., the green star).

---

By expanding Eq. (2) and discarding the constant term $y^{\mathrm{T}}y$, since the constant term does not shift the position of the minimum, we obtain the equivalent formulation:

$$\hat{x} = \underset{x\geq 0}{\operatorname{argmin}}[x^{\mathrm{T}}(A^{\mathrm{T}}A)x - 2(A^{\mathrm{T}}y)^{\mathrm{T}}x]. \tag{3}$$

For a binary target object, $x$ degenerates to the binary variable $\sigma \in \{0,1\}^{\mathrm{N}}$, and Eq. (3) simplifies to Eq. (4), which corresponds to the standard form of quadratic unconstrained binary optimization (QUBO) problem with linear terms $\hat{\sigma} = \operatorname{argmin}_{\sigma}(\sigma^{\mathrm{T}}Q\sigma + 2c^{\mathrm{T}}\sigma)$, where the quadratic coefficient matrix is $Q = A^{\mathrm{T}}A$ and the linear coefficient vector is $c = -2(A^{\mathrm{T}}y)^{\mathrm{T}}$. For grayscale objects, $x$ should be decomposed into a set of binary variables which can also be rearranged into the standard QUBO form in Eq. (4). This work aims to propose a computational compressive reconstruction framework driven by an Ising machine, therefore, we adopt binary objects to simplify the reconstruction demonstrations of single-pixel imaging and diffuser camera experiments.

$$\hat{\sigma} = \underset{\sigma\in\{0,1\}}{\operatorname{argmin}}[\sigma^{\mathrm{T}}(A^{\mathrm{T}}A)\sigma - 2(A^{\mathrm{T}}y)^{\mathrm{T}}\sigma]. \tag{4}$$

We further map $\sigma_{\mathrm{i}}$ to binary spin $s_i \in \{-1,1\}$ via the following transformation:

$$\sigma_{\mathrm{i}} = \frac{1+s_{\mathrm{i}}}{2}. \tag{5}$$

Equation (4) is transformed according to Eq. (5) into:

$$\hat{s} = \underset{s\in\{-1,1\}}{\operatorname{argmin}}\left[\left(\frac{I+s}{2}\right)^{\mathrm{T}} A^{\mathrm{T}}A\left(\frac{I+s}{2}\right) - 2(A^{\mathrm{T}}y)^{\mathrm{T}}\left(\frac{I+s}{2}\right)\right]$$
$$= \underset{s\in\{-1,1\}}{\operatorname{argmin}}\left[\left(\tfrac{1}{4}s^{\mathrm{T}}A^{\mathrm{T}}As + \tfrac{1}{2}I^{\mathrm{T}}A^{\mathrm{T}}As + \tfrac{1}{4}I^{\mathrm{T}}A^{\mathrm{T}}AI\right) - (A^{\mathrm{T}}y)^{\mathrm{T}}I - (A^{\mathrm{T}}y)^{\mathrm{T}}s\right], \tag{6}$$

where $I$ is an N×1 all-ones vector. After discarding the constant term $\left[\frac{1}{4}I^{\mathrm{T}}A^{\mathrm{T}}AI - (A^{\mathrm{T}}y)^{\mathrm{T}}I\right]$ and rearrangement Eq. (6), we obtain Eq. (7):

$$\hat{s} = \underset{s\in\{-1,1\}}{\operatorname{argmin}}\left[\tfrac{1}{4}s^{\mathrm{T}}A^{\mathrm{T}}As + \tfrac{1}{2}(A^{\mathrm{T}}AI - 2A^{\mathrm{T}}y)^{\mathrm{T}}s\right]. \tag{7}$$

Element-wise expansion of Eq. (7) into summation yields:

$$\hat{s} = \underset{s\in\{-1,1\}}{\operatorname{argmin}}\left\{-\sum_{1\le i<j\le N}\left(-\tfrac{1}{2}\sum_{k=1}^{M}A_{ki}A_{kj}\right)s_i s_j - \sum_{i=1}^{N} -\tfrac{1}{2}\left(\sum_{k=1}^{M}A_{ki}\sum_{j=1}^{N}A_{kj} - 2\sum_{k=1}^{M}A_{ki}Y_k\right)s_i\right\}, \tag{8}$$

where $k \in \{1,2,3,\cdots,M\}$ represents the k-th measurement (i.e., the raw index of $A$), and $i,j \in \{1,2,3,\cdots,N\}$ represent the pixels' index of the image (i.e., the column indexes of $A$). The summation of diagonal elements of $A^{\mathrm{T}}A$ is a constant term (i.e., $\sum_{i=1}^{N}\sum_{k=1}^{M}A_{ki}^2$) and has been discarded during expansion. Equation (8) is equivalent to the standard Ising model with the coefficient definitions given below:

$$\begin{cases} J_{ij} = -\tfrac{1}{2}\sum_{k=1}^{M}A_{ki}A_{kj} \\ h_i = -\tfrac{1}{2}\left(\sum_{k=1}^{M}A_{ki}\sum_{j=1}^{N}A_{kj} - 2\sum_{k=1}^{M}A_{ki}Y_k\right) \end{cases}. \tag{9}$$

The diagonal elements $J_{ii}$ contribute only a constant and may be set to zero without loss of generality. The image reconstruction task can be modeled as an Ising formulation shown in Eq. (9), followed by image recovery via optical Ising machine evolution, as illustrated in Fig. 1(b). We name this framework of Eq. (9) as Ising-SA. To improve the reconstruction quality of the algorithm, we further incorporate a foreground-sparsity constraint and a smoothness term into Eq. (4), as shown below:

$$\hat{\sigma} = \underset{\sigma\in\{0,1\}}{\operatorname{argmin}}\left[\sigma^{\mathrm{T}}(A^{\mathrm{T}}A)\sigma - 2(A^{\mathrm{T}}y)^{\mathrm{T}}\sigma + \lambda_1\mathcal{L}_{sparse} + \lambda_2\mathcal{L}_{smooth}\right], \tag{10}$$

where $\lambda_1$ and $\lambda_2$ are the corresponding coefficients, and

$$\begin{cases} \mathcal{L}_{sparse} = \|\sigma\|_1 = \sum_{i=1}^{N}\sigma_i \\ \mathcal{L}_{smooth} = \sum_{1\le i<j\le N} d_{ij}\left\|\sigma_i - \sigma_j\right\|_2^2 = \sum_{1\le i<j\le N} d_{ij}(\sigma_i + \sigma_j - 2\sigma_i\sigma_j) \end{cases}. \tag{11}$$

Given that $\sigma_i \in \{0,1\}$, there has $\sigma_i^2 = \sigma_i$. The $d_{ij}$ is the element in the i-th row and j-th column of the pixel adjacency matrix $D \in \mathbb{R}^{N\times N}$, which is pre-generated according to the specific problem and defined as:

$$d_{ij} = \begin{cases} 1, \text{when } \sigma_i \text{ and } \sigma_j \text{ are neighbors} \\ 0, \text{else} \end{cases}. \tag{12}$$

We can also define the degree $\deg(\sigma_i)$ of the i-th pixel $\sigma_i$ as the number of its neighbors, which can be pre-calculated:

$$\deg(\sigma_i) = \sum_{j=1}^{N} d_{ij}. \tag{13}$$

We consider nearest-neighbor connections for each specific pixel, where the four adjacent pixels (up, down, left, and right) are defined as its neighbors, with $\deg(\sigma_i)$ set to 4. Then, we simplify the smoothness term as:

$$\mathcal{L}_{smoot} = \sum_{1\le i<j\le N} -2d_{ij}\sigma_i\sigma_j + \sum_{i=1}^{N}\deg(\sigma_i)\,\sigma_i. \tag{14}$$

Combining the transformation in Eq. (5), we can obtain the updated coefficients for the Ising machine:

$$\begin{cases} J_{ij} = -\tfrac{1}{2}\sum_{k=1}^{M}A_{ki}A_{kj} + \tfrac{\lambda_2}{2}d_{ij} \\ h_i = -\tfrac{1}{2}\left(\sum_{k=1}^{M}A_{ki}\sum_{j=1}^{N}A_{kj} - 2\sum_{k=1}^{M}A_{ki}Y_k\right) - \tfrac{\lambda_1}{2} \end{cases}. \tag{15}$$

The additional constant terms do not affect the search for the ground-state spin configuration. We name the improved framework of Eq. (15) as Ising-SA-D. Since the elements of $J \in \mathbb{R}^{N\times N}$ and $h \in \mathbb{R}^N$ require high-precision encoding. The 8-bit encoding precision of currently available Ising machines leads to ill-conditioning and invalidates the equivalence with the original problem. Hence, we solve the problem using a simulated-annealing Ising solver with double-precision floating-point arithmetic on an electronic computer. Nevertheless, the modulation and encoding precision of emerging Ising machines is continuously improving, which is promising to combine with our approach to realize high-efficient image reconstruction in the future.

**Simulations**

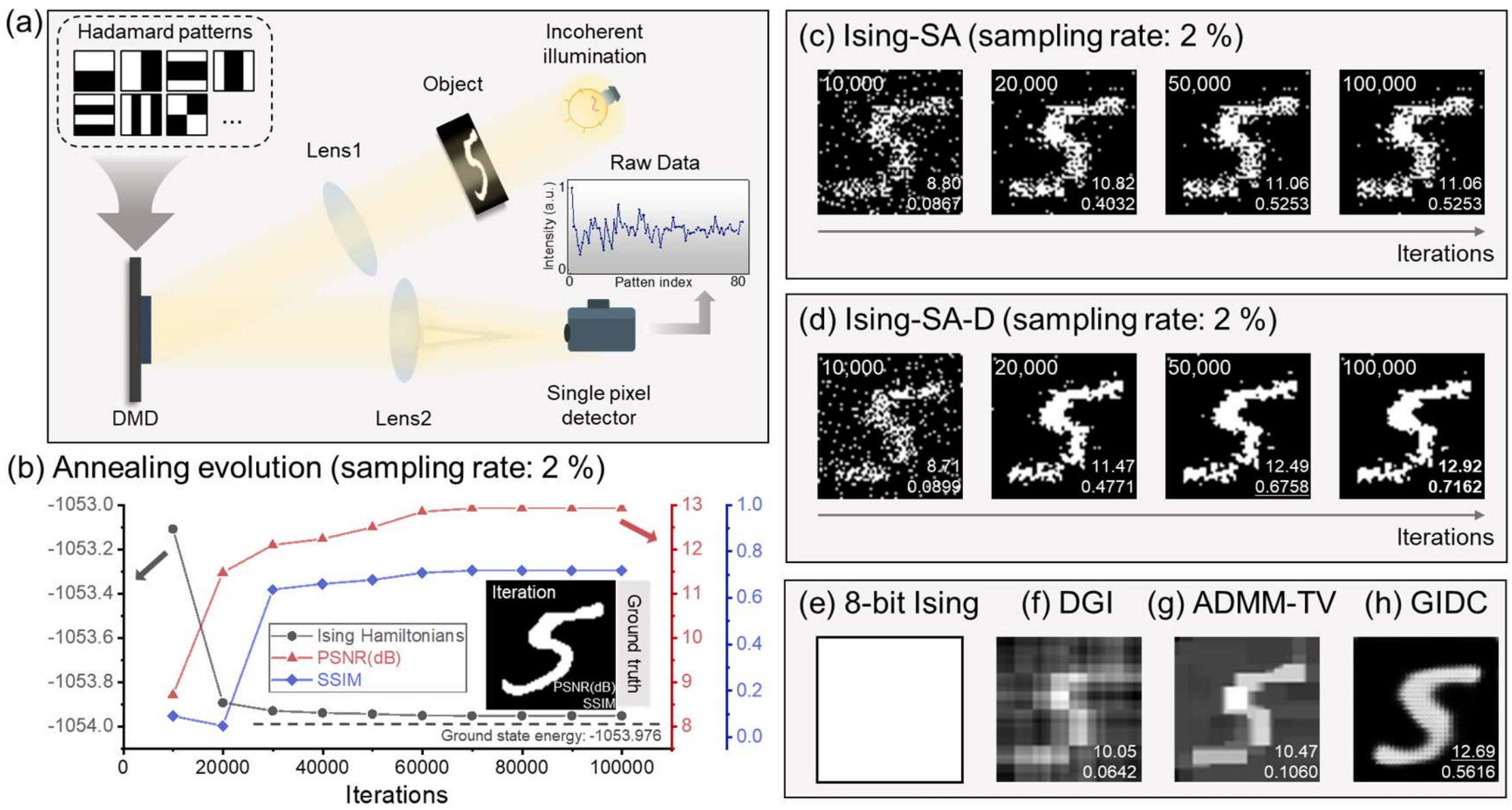

**Figure 2.** The simulation results of single-pixel imaging with different reconstruction methods. (a) The experimental setup of single-pixel imaging using the cake-cutting ordered Hadamard patterns, which is consistent with the simulation setup. The target object is the handwritten digit "5", and the sampling rate in this simulation is 2 %. (b) The annealing evolution of the Ising-SA-D framework, including the Ising Hamiltonian (in gray), PSNR (in red), and SSIM (in blue) of the reconstructed images varying with iterations. The ground truth image is inserted in (b). (c) The reconstructed images recovered by the Ising-SA reconstruction framework at 10,000, 20,000, 50,000, and 100,000 iterations. (d) The reconstructed images recovered by the Ising-SA-D reconstruction framework at 10,000, 20,000, 50,000, and 100,000 iterations. The corresponding iteration numbers are displayed at the top-left corner of each reconstructed image in (c–d). (e) The reconstruction image from the employed 8-bit commercial Ising machine. (f–h) The reconstruction images recovered by DGI, ADMM-TV, and GIDC, respectively. The corresponding PSNR and SSIM values are shown at the bottom-left corner of each reconstructed image in (c–d, f–h). Among these values, the highest values are bolded, and the second-highest are underlined, respectively. The resolution of the reconstructed images is 64×64 pixels.

---

We demonstrate the results of the single-pixel imaging simulations at a sampling rate of 2 % with our proposed frameworks (i.e., Ising-SA, and Ising-SA-D), commercial Ising machine with 8-bit encoding precision (i.e., 8-bit Ising) [56], differential ghost imaging (DGI) [58], alternating direction method of multipliers with total variation regularization (ADMM-TV) [59, 60], and ghost imaging using deep neural-network constraint (GIDC) [61] in Fig. 2. As shown in Fig. 2(b), the Ising Hamiltonian of Ising-SA-D framework decreases with iterations, while the PSNR and SSIM of the corresponding reconstructed images gradually increase. It should be noted that setting the iteration range from 0 to 100,000 does not represent the time-consuming bottleneck of our framework. Instead, this setup is intended to acquire a relatively optimal solution via the simulated annealing (SA) algorithm, for which slower annealing leads to results closer to the ground truth [62-64]. This allows us to estimate the reconstruction performance achievable by physical hardware-based Ising machines in the future more accurately, since many existing ultrafast Ising machines have been demonstrated to find superior solutions compared with the SA algorithm [35-36, 42, 55].

Figure 2(c-d) also demonstrate that the quality of reconstructed images improves with increasing iterations visually. Ising-SA-D framework incorporating sparse and smooth constraints achieves higher reconstruction quality than the Ising-SA framework with only the data-fidelity term at the same iteration, which implies the potential advantages of introducing additional constraints and regularization terms into the Ising-SA framework for other scenarios and applications. Figure 2(e) shows the reconstruction result of the commercial Ising machine, which yields an all-white image (i.e., an all-ones solution vector), and obviously fails to match the ground-truth image. This demonstrates that the 8-bit precision is insufficient to properly encode this compressive image reconstruction problem. Figures 2 (f–h) present the reconstruction images of three mainstream algorithms under the same sampling rate of 2%, whose metrics are inferior to those reconstructed by the Ising-SA-D framework at 100,000 iterations.

**Experiments**

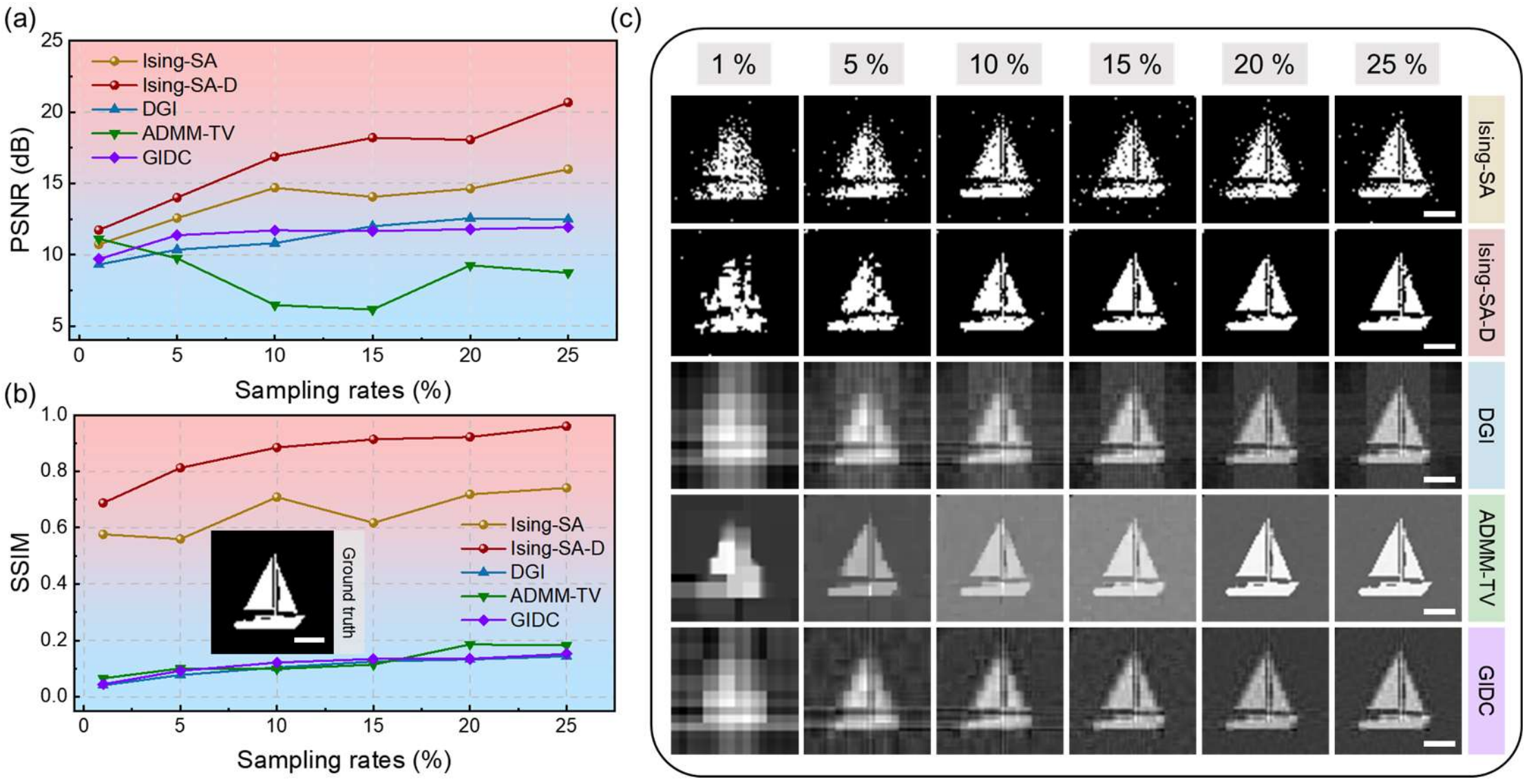

**Figure 3.** The results of single-pixel imaging experiments with different reconstruction methods. (a), (b) The PSNRs and SSIMs of Ising-SA (in brown), Ising-SA-D (in red), DGI (in blue), ADMM-TV (in green), and GIDC (in purple) at different sampling rates (1 %, 5 %, 10 %, 15 %, 20 %, and 25 %) are shown, respectively. The ground truth of the employed object "sailboat" is inserted in (b). (c) The corresponding reconstructed images of different methods at different sampling rates. The resolution of the reconstructed images is 64×64 pixels, and the scale bar is 0.5 cm.

---

Figure 3 demonstrates the reconstructed results of single-pixel imaging experiments. The experimental setup is consistent with Fig. 2(a). A parallel coaxial LED light (LB-CLS-50, LBTEK) illuminates an amplitude-type spatial light modulator (A-SLM, TSLM023-A, Cas Microstar), which is employed to project the object. The object is imaged to the plane of the digital micro-mirror device (DMD, V-7001-VIS, ViALUX) via a lens (OLC320754M, JCOPTiX). The cake-cutting ordered Hadamard patterns are projected on the DMD. The reflected light from the DMD is focused by a condenser lens (OLS300434-T2M, JCOPTiX), followed by integrating measurement with a single-pixel detector (PDAPC2, Thorlabs) after switching. A data acquisition device (USB-3133A, ART Technology) is utilized for signal acquisition. The target object images are custom-designed in-house.

The Ising-SA-D framework achieves the best image reconstruction quality at each sampling rate, followed by the Ising-SA framework, with both set for 100,000 iterations. The reconstruction quality of all methods improves as the sampling rates increase. Although mainstream algorithms can visually recover the target objects, they exhibit high-level background noise, yielding lower SNR and metric values (PSNRs and SSIMs) compared with our proposed frameworks. We also conduct the same experimental acquisition and reconstruction on other target objects to demonstrate the generalization and effectiveness of our approach, and the corresponding results are presented in **Appendix A**.

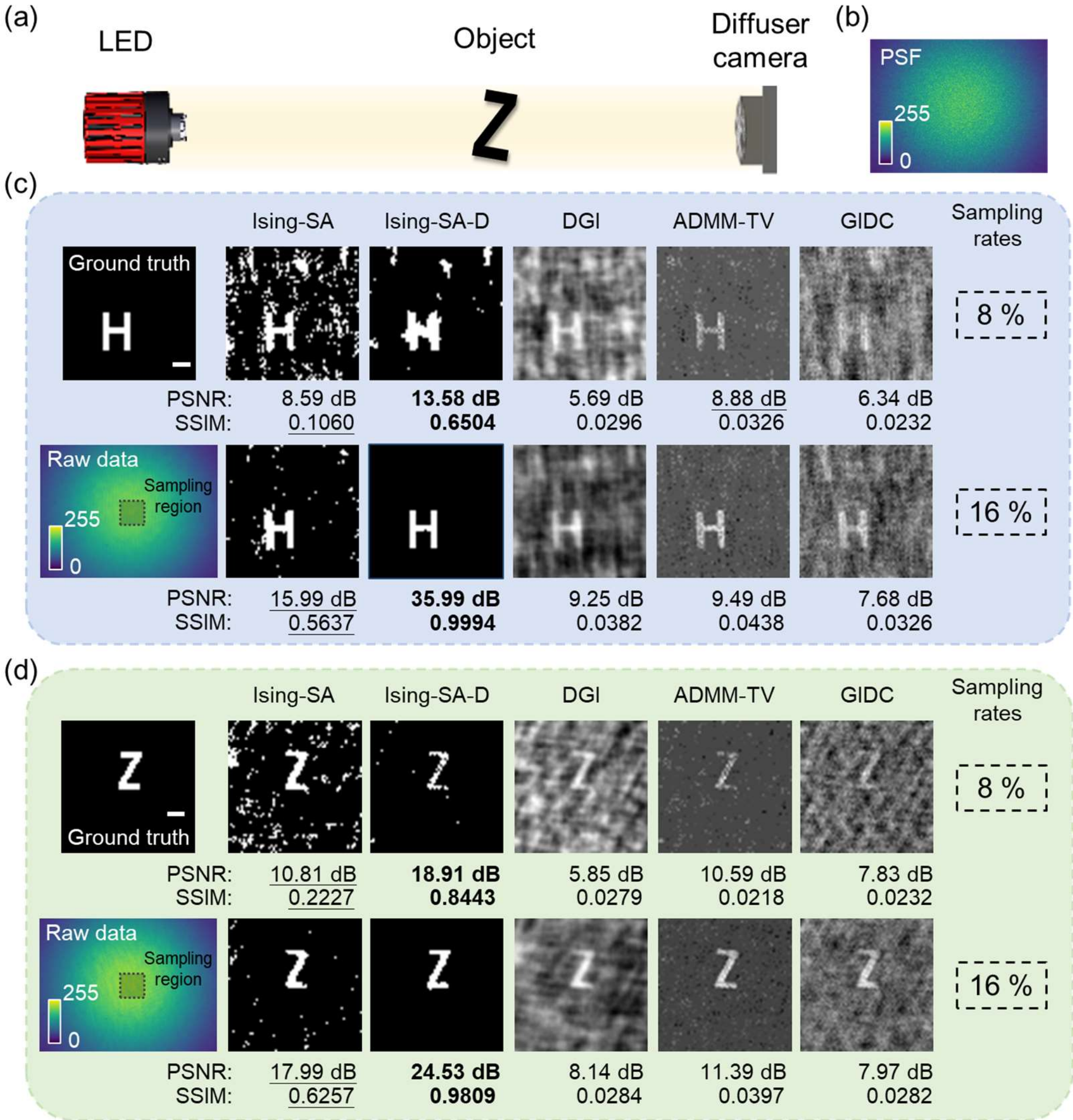

**Figure 4.** The results of the experiments of the diffuser camera. (a) The experimental setup. An LED light illuminates the structured object, followed by capture by the custom-designed diffuser camera. (b) The pre-calibrated PSF of the system at the object position. (c-d) The captured raw data, the reconstructed images, and the metric values (PSNRs and SSIMs) of the 3D-printed transmitted objects "H" and "Z" with different methods at the sampling rates of 8 % and 16 %, respectively. The sampling region within each raw data is marked by a dashed box. This square region is centered at the raw-data center, and its size varies with different sampling rates. The highest metric values for each target object at each sampling rate are bolded, and the second-highest values are underlined. The resolution of the reconstructed images is 64×64 pixels, and the scale bar is 1 cm.

We also applied the proposed framework to inverse reconstruct the images of diffuser camera experiments, as shown in Fig. 4. Figure 4(a) presents the experimental setup. A monochromatic red LED light (GCI-060401, DHC), which is collimated by a lens (MCX10610, LBTEK), illuminates the 3D-printed transmission-structured objects (i.e., "H" and "Z"), which are placed within the range of the optical memory effect. The transmitted light is modulated by a diffuser (GCL-201101, DHC), followed by capture by a two-dimensional sensor (daA2500-14um, Basler) which is placed 1 cm behind. The PSF is pre-calibrated by replacing the object with a 100 μm pinhole (GCO-P100A, DHC) at the same position, as presented in Fig. 4(b).

The vectorized measurement $Y$ of the forward model is composed of the values of the sampling region in each raw data, and the measurement matrix $A$ is constructed by the PSF. As shown in Fig. 4(c-d), our proposed Ising-SA and Ising-SA-D frameworks with 100,000 iterations perform better than other methods at low-sampling rates, which implies the application potential of our proposed framework in other computational compressive imaging techniques with limited raw measurements. Although the demonstration is performed numerically with double-precision solvers, advances in hardware encoding precision of physical Ising machines will enable on-chip, low-latency image reconstruction in future work. The presented strategy can also be extended to other inverse problems beyond diffuser-camera and single-pixel imaging.

**Conclusion and Discussion**

In summary, we have proposed and demonstrated the compressive imaging reconstruction framework, which models the reconstruction task derived from the optical forward model, consisting of a data-fidelity term, a sparsity constraint term and a smoothness constraint term, into the Ising model formulation. We have demonstrated the effectiveness of this framework with a simulated-annealing-based Ising solver, and have achieved the higher reconstruction quality than the mainstream algorithms in the simulation and experiments of single-pixel imaging and lensless diffuser camera. More imaging techniques and modalities are potentially inspired to be incorporated into this framework [65-67]. The proposed approach is further going to be deployed to the developing physical Ising machines to achieve the ultrafast and energy-efficient imaging reconstruction.

We should note that, besides high-precision encoding requirements, the proposed method demands fully-connected Ising machine spin resources scaling as $O(N^2)$ and $O(N^4)$ couplings for reconstructing binary images with a resolution of $N\times N$ pixels. For grayscale images with the same resolution, the required spins and couplings scale as $O(aN^2)$ and $O(a^2N^4)$, respectively, where $a$ represents the gray level count ($a$ equals 256 for 8-bit grayscale images). We reconstruct 64×64 pixels binary images with 4096+64=4160 virtual spins in our demonstrations. Large-scale fully-connected Ising machines are difficult to realize on physical platforms including coherent Ising machines [35, 36, 46]. Spatial photonic Ising machines have been proved to possess intrinsic full connectivity [68] and are progressing toward larger size [48, 69], complex couplings [48] and improved precision [70]. Dedicated reconstruction frameworks designed for spatial photonic Ising machines may greatly reduce the hardware resource cost, making this approach a promising route toward rapid real-world implementation.

**Appendix A: Reconstruction of other objects in experimental single-pixel imaging**

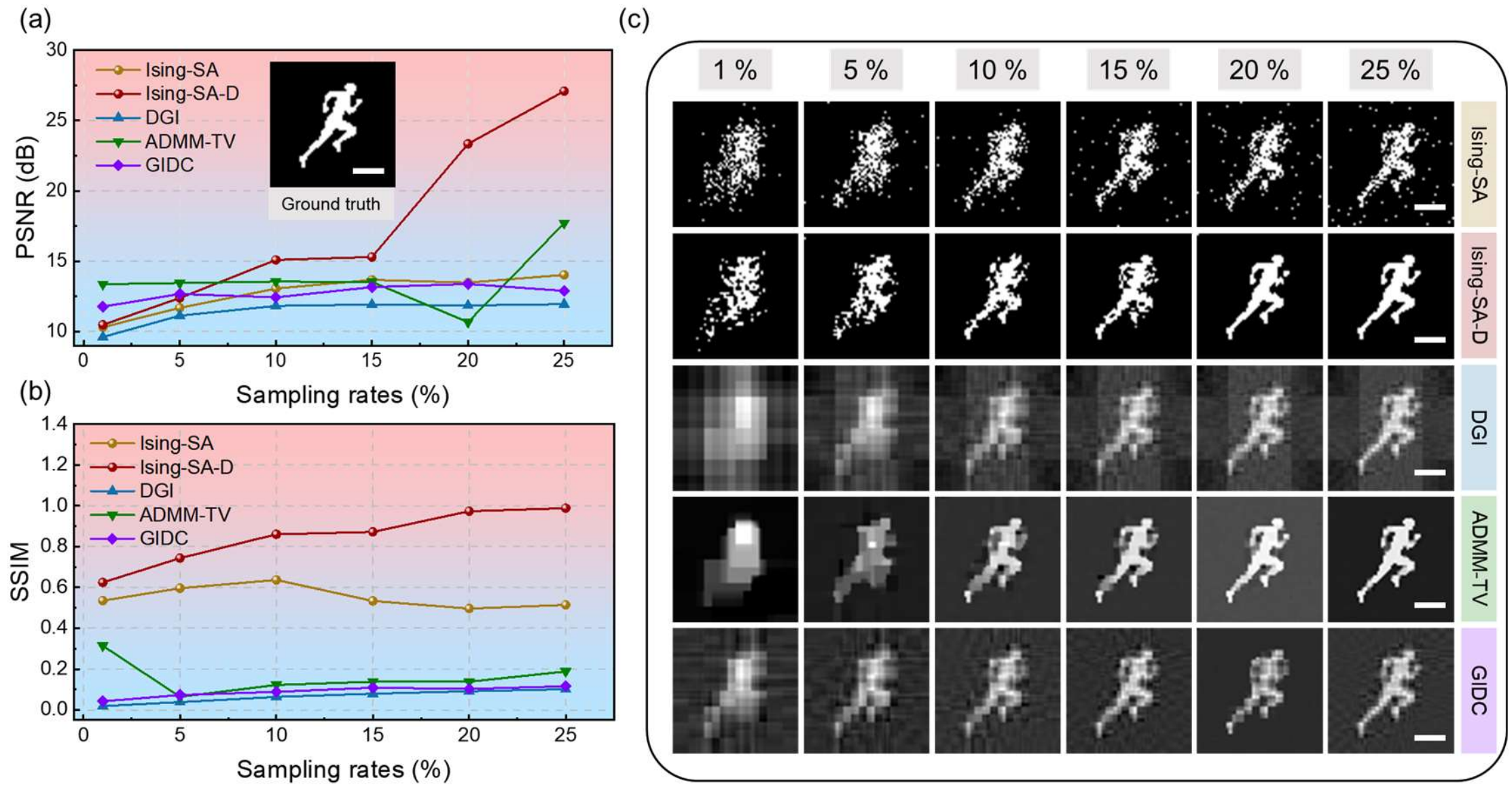

**Figure 5.** The results of single-pixel imaging experiments with different reconstruction methods. (a), (b) The PSNRs and SSIMs of Ising-SA (in brown), Ising-SA-D (in red), DGI (in blue), ADMM-TV (in green), and GIDC (in purple) at different sampling rates (1 %, 5 %, 10 %, 15 %, 20 %, and 25 %) are shown, respectively. The ground truth of the employed object "running man" is inserted in (a). (c) The corresponding reconstructed images of different methods at different sampling rates. The resolution of the reconstructed images is 64×64 pixels, and the scale bar is 0.5 cm.

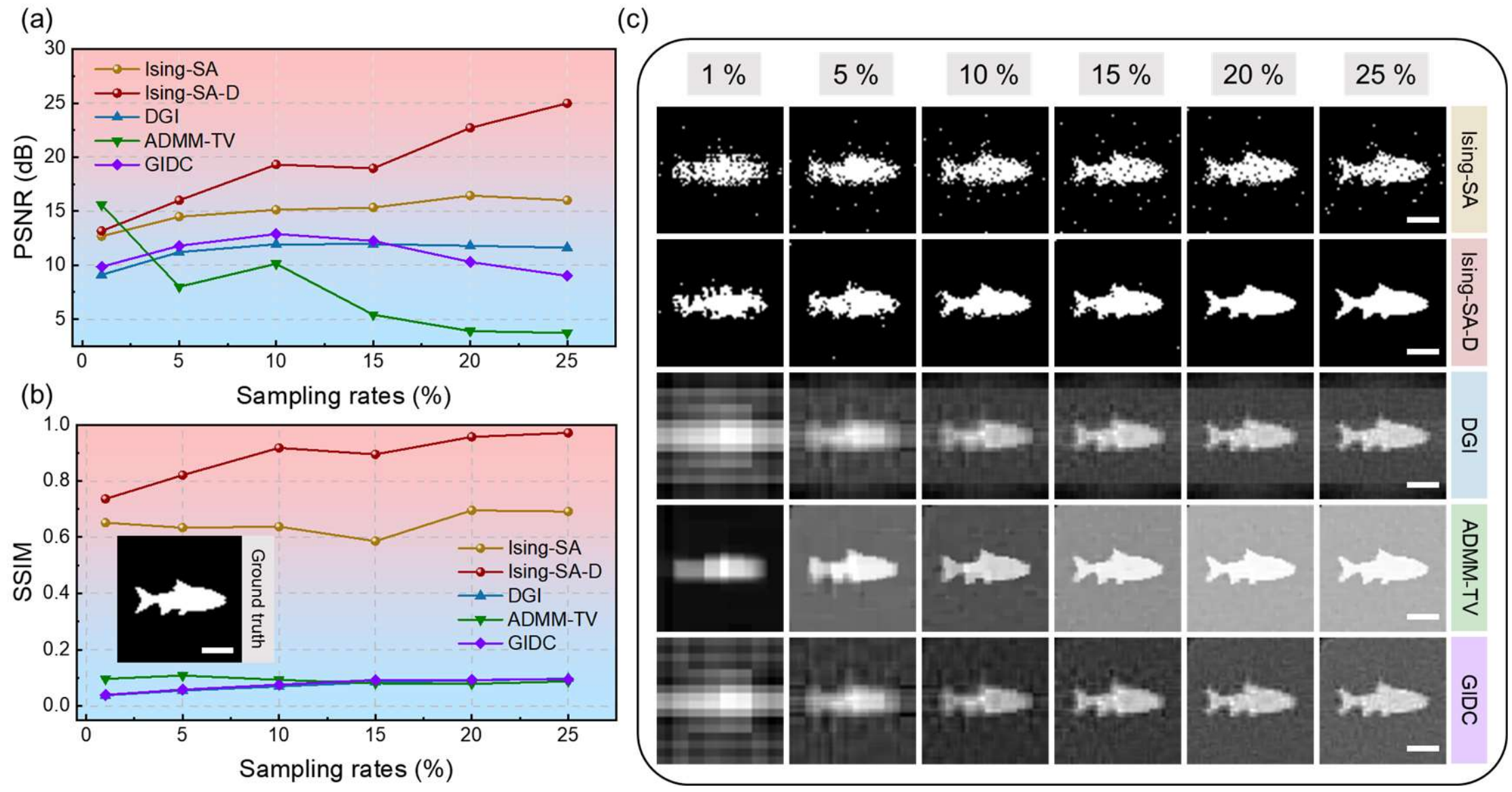

**Figure 6.** The results of single-pixel imaging experiments with different reconstruction methods. (a), (b) The PSNRs and SSIMs of Ising-SA (in brown), Ising-SA-D (in red), DGI (in blue), ADMM-TV (in green), and GIDC (in purple) at different sampling rates (1 %, 5 %, 10 %, 15 %, 20 %, and 25 %) are shown, respectively. The ground truth of the employed object "fish" is inserted in (b). (c) The corresponding reconstructed images of different methods at different sampling rates. The resolution of the reconstructed images is 64×64 pixels, and the scale bar is 0.5 cm.

**Acknowledgments**

This work was supported by the National Key R&D Program of China (No. 2025YFF0515504), the National Natural Science Foundation of China (Nos. 62401359 and 62471289), the State Key Laboratory of Photonics and Communications, the Quantum Science and Technology - National Science and Technology Major Project (No. 2021ZD0300703), the Shanghai Municipal Science and Technology Major Project (No. 2019SHZDZX01), and the CCF-QBoson Fund for Quantum-Computing Application Innovation (No. CCF-QBoson202406).

**Author Contributions**

G.Z. conceived the research project. Z.Z. designed the scheme, implemented the algorithms, developed the simulation, and performed the experiments. G.Z. and T.X. supervised the project. Z.Z. analyzed all the results and prepared the manuscript, with assistance from X.W.. All the authors participated in the discussion and confirmed the final manuscript.

**Conflicts of Interest**

The authors declare no conflicts of interest.

**Data Availability Statement**

The data that support the findings of this study are available from the corresponding author upon reasonable request